\documentclass[pra,twocolumn,english,showpacs]{revtex4-1}
\usepackage{graphicx}
\usepackage{amssymb}
\usepackage[rightcaption]{sidecap}

\begin{document}

\title{Laser spectroscopy and cooling of Yb$^+$ ions on a deep-UV transition}

\author{Hendrik M. Meyer}
\author{Matthias Steiner}
\author{Lothar Ratschbacher}
\author{Christoph Zipkes}
\author{Michael K{\"o}hl}

\affiliation{Cavendish\,Laboratory,\,University~of Cambridge, JJ Thomson Avenue, Cambridge CB30HE, United Kingdom}

\begin{abstract}
We perform  laser spectroscopy of Yb$^+$ ions on the 4f$^{14}$6s $^2$S$_{1/2}$ - 4f$^{13}$5d6s $^3$D$[3/2]_{1/2}$ transition at 297 nm.  The frequency measurements for  $^{170}$Yb$^+$, $^{172}$Yb$^+$, $^{174}$Yb$^+$, and $^{176}$Yb$^+$ reveal the specific mass shift as well as the field shifts. In addition, we demonstrate laser cooling of Yb$^+$ ions using this transition and show that light at 297~nm  can be used as the second step in the photoionization of neutral Yb atoms.

\end{abstract}


\date{\today}

\maketitle

\section{Introduction}

In the last decade trapped Yb\textsuperscript{+} ions have been a workhorse in the fields of quantum information processing \cite{Moehring2007,Kim2010,Timoney2011}, precision measurements \cite{Schneider2005,Roberts2000} and hybrid systems \cite{Zipkes2010}.  Yb$^+$ has a rich energy level structure because besides pure valence-electron excitations, also electrons from a closed f-shell can be excited. This leads to unusual features such as the electric octupole transition $^2$S$_ {1/2}$-$^2$F$_{7/2}$, which is currently explored for the prospect of  an ion based atomic clock, as it has a  natural lifetime of several years \cite{Roberts2000,Roberts1997}.

\begin{figure}
 \includegraphics[width=.9\columnwidth,clip=true]{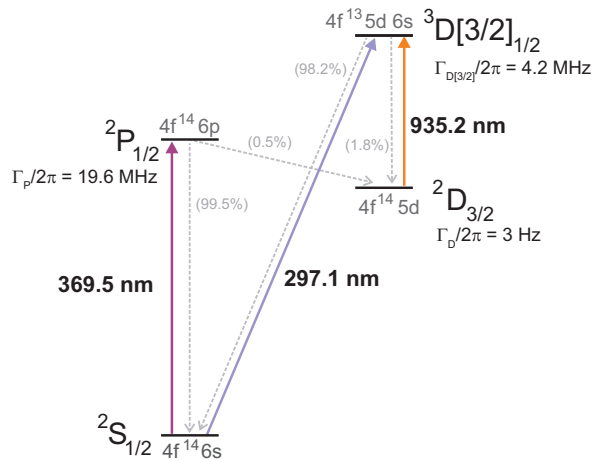}%
\caption{(color online) Relevant energy levels and transitions of Yb$^+$ (branching ratios given in percentages). $\Gamma_{D[3/2]}$, $\Gamma_{D}$ and branching ratio out of the $^2$P$_{1/2}$ state are taken from Ref.~\cite{PhysRevA.76.052314},  $\Gamma_{D[3/2]}$ from Ref.~\cite{Berends1993} and the $^3$D$[3/2]_{1/2}$ branching ratio from Ref.~\cite{Biemont1998}.}
\label{ybion_terms_spec}
 \end{figure}

Since the work of \cite{PhysRevA.44.R20}, a common feature of most trapped Yb\textsuperscript{+} experiments is the continuous excitation of the ion within a four level system (Figure \ref{ybion_terms_spec}). The ion  is cooled and detected on the $^2$S$_ {1/2}$-$^2$P$_{1/2}$ transition near 369 nm. A second decay channel to the metastable $^2$D$_{3/2}$ state necessitates a second laser to avoid interruptions of the fluorescence. Light near 935 nm repopulates the ground state via the $^3$D$[3/2]_{1/2}$~state.

In this paper, we extend the available laser manipulation toolbox for trapped Yb\textsuperscript{+} ions by exciting the 4f$^{14}$6s~$^2$S$_{1/2}$~-~4f$^{13}$5d6s~$^3$D$[3/2]_{1/2}$ transition at 297 nm. We perform laser spectroscopy and present frequency measurements for four isotopes. The isotope shifts contain information about the change of the charge distribution within the nucleus and about the correlations of electrons, which are intricate because of a partially filled f-shell in Yb\textsuperscript{+}. Therefore they can be used to test ab-initio calculations.  In addition, we show that the 297 nm light can be used as second the step in the photoionization process of neutral Yb as well as for laser cooling of single Yb\textsuperscript{+} ions. The relatively narrow line width could potentially allow to reach lower laser cooling temperatures.

\section{Experimental Setup}

We trap single Yb\textsuperscript{+} ions in a RF Paul trap, which consists of two opposing endcap electrodes \cite{Schrama1993} separated by 150 $\mu$m. The electrodes are formed from tungsten wire of 250 $\mu$m diameter and are etched to a needle-like geometry \cite{PhysRevLett.97.103007}. By applying a radio frequency signal of 20 MHz and a few hundred Volts amplitude, the ion is trapped with secular trap frequencies of $\omega_{ax}/2\pi\approx 3.7$~MHz and $\omega_{rad}/2\pi\approx  1.7$~MHz.
Single ions are loaded into the trap by  two-step photoionization from a thermal atomic beam \cite{PhysRevA.73.041407}. A stainless steel tube filled with Yb is resistively heated by a short current pulse (typically 60-80~ms, 60~A) and emits neutral Yb atoms into the trap. The atoms are resonantly excited on the $^1$S$_{0}$-$^1$P$_{1}$ transition with laser light at 398~nm and subsequently ionized by a laser at 369~nm. The first excitation step is isotope selective. After ionization, the ion is continuously laser cooled on the  $^2$S$_ {1/2}$-$^2$P$_{1/2}$ transition and the resulting fluorescence is collected by an in-vacuum objective and detected by a photomultiplier tube.

\begin{figure}
 \includegraphics[width=\columnwidth,clip=true]{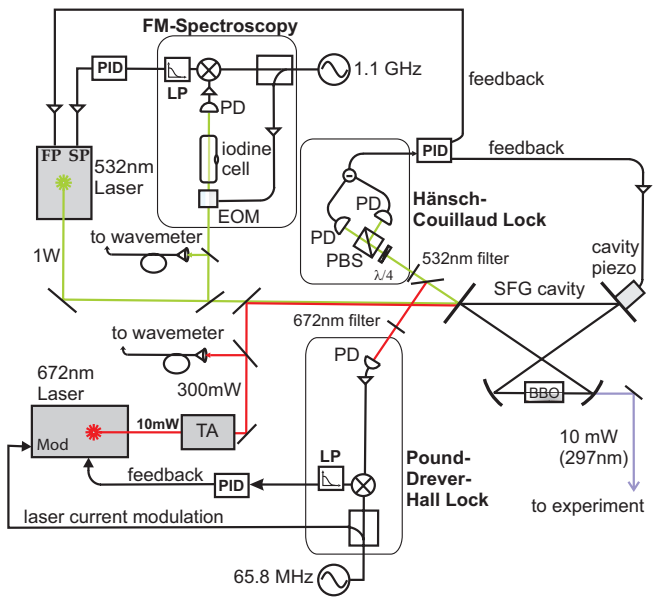}%
\caption{(color online) Laser system to generate 297 nm light. Abbreviations: EOM: Electro-optic modulator, FP: Fast piezo-transducer, LP: Low pass, Mod: Laser current modulation, PBS: Polarizing beam splitter, PD: Photodiode,  PID: Feedback controller, SFG: Sum frequency  generation, SP: Slow piezo-transducer, TA: Tapered Amplifier}
\label{green_setup}
 \end{figure}

The laser system for exciting the $^2$S$_{1/2}$-$^3$D$[3/2]_{1/2}$ transition at 297~nm is based on sum frequency  generation (SFG) of light at 532~nm and 672~nm. The sketch of the optical setup is shown in Figure \ref{green_setup}. The green 532~nm laser is a commercial system (Coherent Verdi~V6), whereas the red 672~nm diode laser with subsequent amplification stage is homebuilt.  In order to supply an atomic frequency reference, the green laser is stabilized to the R(56)32-0 transition in molecular Iodine \cite{Jungner1995}. As the study of isotope shifts in Yb\textsuperscript{+} requires a large frequency scan range, we use the Doppler broadened absorption line (at room temperature: line width $\approx 0.8$~GHz) and generate an error signal by FM spectroscopy \cite{Bjorklund1983}. We measured the rms frequency fluctuations of the green laser to be about 20~MHz.

We achieve type-I (ooe) SFG in a doubly resonant bow-tie cavity (finesse: 75, free spectral range: $1.07$~GHz). The cavity length is locked to the 532~nm laser using the H{\"a}nsch-Couillaud stabilization technique \cite{Hansch1980441}, whereas the 672~nm laser follows the SFG cavity by the Pound-Drever-Hall locking technique \cite{Drever1983}. By sweeping the green 532~nm laser we can scan the generated UV light continuously about 1~GHz. Input light intensities of 1~W~(532~nm) and 300~mW (672~nm) yield about 10~mW of 297~nm light.

\section{Isotope Shifts}

\begin{figure}
 \includegraphics[width=.9\columnwidth,clip=true]{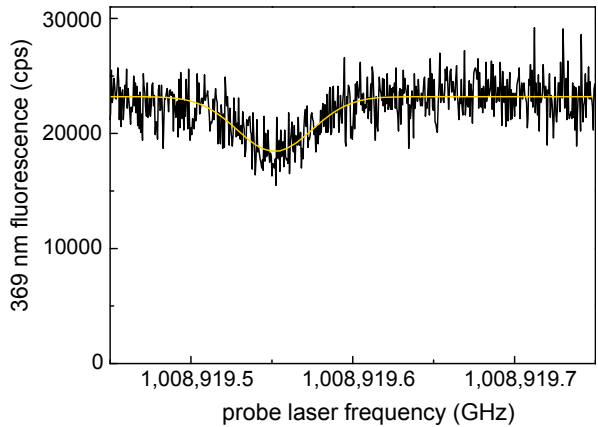}%
\caption{(color online) Double-resonance spectrum of the $^2$S$_{1/2}$-$^3$D$[3/2]_{1/2}$ transition in $^{176}$Yb\textsuperscript{+}. Yellow line: Gauss fit. Transition frequencies have been determined by averaging 10 spectra.}
\label{typicalSpec}
 \end{figure}

Spectroscopy on the \mbox{$^2$S$_{1/2}$-$^3$D$[3/2]_{1/2}$} transition is performed in a double-resonance scheme. The laser at 369~nm cools the ion continuously on the \mbox{$^2$S$_ {1/2}$-$^2$P$_{1/2}$} transition and yields fluorescence, which is detected. Simultaneously, the 297~nm probe laser excites the \mbox{$^2$S$_{1/2}$-$^3$D$[3/2]_{1/2}$} transition. Additional 935~nm light is necessary to deplete the population of the \mbox{$^3$D$[3/2]_{1/2}$} state. By sweeping the frequency of the 297~nm light, the \mbox{$^2$S$_{1/2}$-$^3$D$[3/2]_{1/2}$} resonance is observed as a drop in the 369 nm fluorescence (Figure \ref{typicalSpec}).

The ratio of the saturation parameters, $s_{369nm}$ and $s_{297nm}$ ($s_\lambda = \frac{I_\lambda}{I_{\lambda, sat}}$, with the (saturation) intensity $I_{\lambda}$ ($I_{\lambda, sat}$)) is important for the strength of the observed signal. For  $s_{369nm} \gg s_{297nm}$ the excitation into the $^3$D$[3/2]_{1/2}$ is much slower than the probing of the $^2$S$_{1/2}$ state due to the 369~nm laser, which effectively suppresses the transition to  $^3$D$[3/2]_{1/2}$ in a quantum zeno like manner. In contrast, the signal of the 369~nm $^2$S$_ {1/2}$-$^2$P$_{1/2}$ fluorescence shows a significant drop for similar saturation parameters $s_{369nm}~\approx~s_{297nm}$. The steady state fluorescence is proportional to the population of the $^2$P$_{1/2}$ state, which is given by

\begin{equation}
p_{P}  =  \frac{1}{2} \cdot  \frac{s_{369nm}}{1 + s_{369nm} + \left( \frac {2 \Delta_{369nm}}{ \Gamma_P} \right)^2 + \epsilon} ,
\end{equation}
assuming fast repumping out of the $^2$D$_{3/2}$ state and neglecting coherences. Here  $\Delta_{369nm}$ is the detuning of the 369~nm laser and $\Gamma_P$ the natural line width of the $^2$P$_{1/2}$ state. The term $\epsilon$ describes the coupling to the $^3$D$[3/2]_{1/2}$ state
\begin{equation}
\epsilon =  \frac{1}{2} \cdot \frac{s_{297nm} \left(  2 + s_{369nm} + 2  \left( \frac {2 \Delta_{369nm}}{ \Gamma_P} \right)^2 \right)} {2 + s_{297nm} + \left( 2 \frac {2 \Delta_{297nm}}{ \Gamma_{D[3/2]}} \right)^2 } ,
\end{equation}
where $\Delta_{297nm}$ is the detuning of the 297~nm laser  and  $\Gamma_{D[3/2]}$ the natural line width of the $^3$D$[3/2]_{1/2}$ state. For typical experimental parameters of $s_{369nm}~\approx~s_{297nm}~\approx~2-5$ and $\Delta_{369nm}~\approx~-\Gamma_{P}/2 $, this results in a 20$\%$-30\% drop of the fluorescence, which is consistent with our measured spectra. We typically scan the 297~nm light 400~MHz in 10~s and record the 369~nm fluorescence. The line width of the 297~nm laser, residual micromotion and the continuous driving of the \mbox{$^2$S$_ {1/2}$-$^2$P$_{1/2}$} transition broaden the linewidth, which is measured to be 40~MHz. The absolute frequency is determined by measuring the frequency of the 532~nm and the 672~nm laser with a wave meter (HighFinesse~WS/07). The wave meter is calibrated to the D$_2$ (F$=2$, F'$=(2,3)$) crossover line in $^{87}$Rb and has a specified 3-$\sigma$ accuracy of 60~MHz. In this manner we measured the \mbox{$^2$S$_{1/2}$-$^3$D$[3/2]_{1/2}$}  frequencies for the isotopes $^{170}$Yb$^+$, $^{172}$Yb$^+$, $^{174}$Yb\textsuperscript{+} and $^{176}$Yb\textsuperscript{+} (Table \ref{tableFreq}).

\begin{table}
 \caption{Measured frequencies of the $^2$S$_{1/2}$-$^3$D$[3/2]_{1/2}$ transition in Yb$^+$ (1-$\sigma$ error).}
 \label{tableFreq}
\begin{ruledtabular}
 \begin{tabular}{ll}
 Isotope & Frequency (THz) \\ \hline
\textsuperscript{170}Yb$^+$ & 1008.91619(3)\\
\textsuperscript{172}Yb$^+$ & 1008.91759(3)\\
\textsuperscript{174}Yb$^+$ & 1008.91855(3)\\
\textsuperscript{176}Yb$^+$ & 1008.91958(3)\\
 \end{tabular}
\end{ruledtabular}
 \end{table}

The isotope shift of a spectral line consists of two contributions, the field and the mass shift \cite{King}. The field~shift~(FS) is caused by the change of the charge distribution within the nucleus and is very sensitive to the electron charge density at the nucleus $|\Psi(0)|^2$. The mass shift itself has  two contributions: The change in the reduced mass is taken into account by the normal~mass~shift~(NMS) whereas the specific~mass~shift~(SMS) describes the change in the correlations between electrons. In total, the isotope shift of a spectral line $\alpha$ between isotopes a and b reads
\begin{equation}
\Delta \nu^{\alpha}_{a,b} =   \delta \nu^{\alpha}_{FS,a,b} + \left(  k^{\alpha}_{NMS} + k^{\alpha}_{SMS} \right)   \frac{   M_b - M_a  } {M_a   \left(  M_b + m_e \right)  } ,
\end{equation}
with $M_a$ and $M_b$ ($M_a < M_b$) being the isotope masses and $m_e$ the mass of the electron. The parameter $k^{\alpha}_{NMS}$ depends only on the transition frequency of the lighter isotope $\nu_a$  and is given by  $k^{\alpha}_{NMS}  = \nu^{\alpha}_a m_e$. A common technique to identify the field shift and the specific mass shift from measured isotope shifts $\Delta \nu^{\alpha}_{a,b}$ is via a King plot \cite{King}. After eliminating the normal mass shift from the measured isotope shift, the residual shift is multiplied by $ \frac {M_a   \left(  M_b + m_e \right)  } {    M_b - M_a  }$ and the reduced mass of a reference pair (we choose $^{170}$Yb,$^{174}$Yb) to obtain the modified isotope shift $\Delta \widetilde{\nu}^{\alpha}_{a,b}$:
\begin{eqnarray}
\Delta  \widetilde{ \nu}^{\alpha}_{a,b} =& \frac{ M_{174} - M_{170} } {M_{170}   \left(  M_{174} + m_e \right)  } \times \nonumber\\
&\left( \frac{M_a   \left(  M_b + m_e \right)} {  M_b - M_a  } \Delta \nu^{\alpha}_{a,b} - \nu^{\alpha}_a m_e  \right) .
\end{eqnarray}

\begin{figure}
 \includegraphics[width=.85\columnwidth,clip=true]{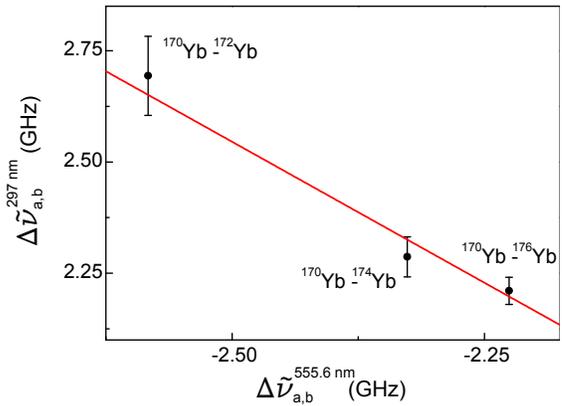}%
\caption{(color online) King plot of our data vs the data of \citet{PhysRevA.20.239}. The slope of the linear fit is used to determine the field shift.}
\label{KingPlot}
 \end{figure}

The modified isotope shift $\widetilde{\nu}^{\alpha}_{a,b}$ is plotted against that of another transition $\beta$ and the slope $ \gamma$ of the resulting linear fit relates the field shifts:
\begin{equation}
\gamma =    \frac {\delta \nu^{\alpha}_{FS,a,b} } { \delta \nu^{\beta}_{FS,a,b}} .
\end{equation}
We use the data from \citet{PhysRevA.20.239} on the 555.6 nm  $^1$S$_0$-$^3$P$_1$ transition in neutral Yb as the reference for the King plot. These data are very precise and allow to calculate the field shift \cite{Zinkstok2002}. For the King plot and the determination of the transition characteristic parameter, $\delta \nu^{\alpha}_{FS,a,b}$ and $k^{\alpha}_{SMS}$, $^{170}$Yb was chosen as reference isotope. From Figure \ref{KingPlot} we extract the slope $\gamma = -1.26 (20)$ and calculate $\delta \nu^{297nm}_{FS,a,b}$  and $k^{297nm}_{SMS}$ (Table  \ref{tableIsotopeShift}).

 \begin{table}
 \caption{Experimentally determined contributions to the isotope shift.}
 \label{tableIsotopeShift}
 \begin{ruledtabular}
 \begin{tabular}{ll }
$\delta \nu^{297nm}_{FS,170,176}$   &  4.16 (66) GHz \\
$\delta \nu^{297nm}_{FS,170,174}$  &  2.92 (47) GHz\\
$\delta \nu^{297nm}_{FS,170,172}$  &  1.65 (26) GHz\\
 $k^{297nm}_{SMS}$         & -4.5 (4.0) GHz amu\\
 \end{tabular}
 \end{ruledtabular}
 \end{table}

Our results can be interpreted in comparison with the data of Zinkstok et al. \cite{Zinkstok2002}, who investigated a similar line in neutral Yb, namely the 4f$^{14}$6s$^2$~$^1$S$_0$ to 4f$^{13}$5d6s$^2$~$^1$P$_1$ transition. They found a similarly large negative specific mass shift, which is explained by the strong coupling of d- and f-electrons \cite{King}. In contrast, our measured field shift is about 50\% higher. This may reflect the fact that, in general the electron charge density at the nucleus is higher in ionic systems than in their neutral counterparts and thereby more sensitive to changes of the charge distribution within the nucleus.

\section{Photoionization and laser cooling}
\begin{figure}
 \includegraphics[width=.9\columnwidth,clip=true]{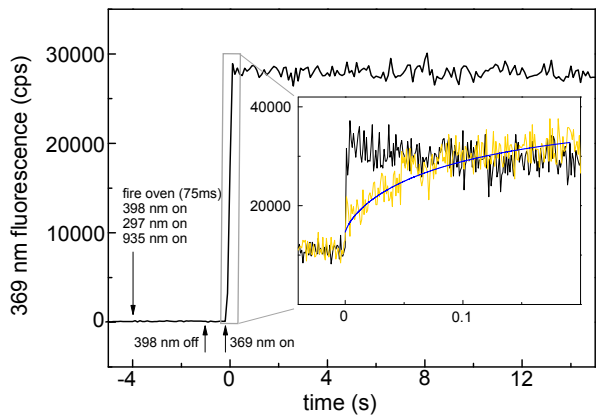}%
\caption{(color online) Recorded 369 nm fluorescence while loading a single Yb\textsuperscript{+} with 297~nm light. Inset: Zoom into the onset of the fluorescence (each curve is the average of 10 loadings). Yellow curve: Large detuning, $\Delta_{297 nm} = -400$~MHz. Blue line: Expected fluorescence of an 100~K hot ion \cite{Wesenberg2007}. Black curve: Sweep of $\Delta_{297 nm}$  from -200~MHz to -60~MHz in 3 s after heating the oven leads to a high initial level of 369 nm fluorescence indicating a cold ion. In both cases $\Delta_{369 nm} = - \Gamma_{P}/2$.}
\label{Loading}
 \end{figure}

Photoionization of Yb can be achieved by a photon at 398~nm ($^1$S$_{0}$-$^1$P$_{1}$ transition) and a second photon with a wavelength smaller than 394~nm. Here we demonstrate that  light at 297~nm can be used for the second step. At the beginning of the loading sequence the Yb oven is heated for 75~ms and simultaneously the 398~nm, the 297~nm and the 935~nm laser lights are switched on. After three seconds the light at 398~nm is turned off and one second later the 369~nm laser illuminates the ion and the resulting fluorescence is recorded (Figure  \ref{Loading}). We note that for reliable loading the oven heating time needs to be slightly longer compared to the standard loading with 369 nm light.

The fluorescence curve after switching on the 369~nm light depends strongly on the kinetic energy of the ion due to the Doppler shift and  can be used to determine its temperature \cite{Wesenberg2007,Epstein2007}. In order to measure the onset of the fluorescence, we interrupt the 935~nm light 10~ms before the 369~nm light is switched on. This allows the 369~nm intensity controller to adjust without affecting the ion. We switch the 935~nm light back on 100~ms later. Only with both lights, 935~nm and 369~nm, the excitation cycle is closed  and the ion continuously scatters  photons. The inset in Figure \ref{Loading} shows the onset of the fluorescence for two different loading sequences. In the first the 297~nm laser is set 400~MHz below the resonance frequency during the loading (yellow curve). Because of the large detuning the ion scatters only few photons on the $^2$S$_{1/2}$-$^3$D$[3/2]_{1/2}$ transition and the cooling of the ion is very inefficient. Thus, the 369~nm fluorescence starts low due to Doppler broadening and increases as the 369~nm laser cools the ion. In the second loading sequence, a smooth shifting of the 297~nm laser frequency  towards the resonance (from $\Delta_{297 nm} = -200$~MHz to $\Delta_{297 nm} = -60$~MHz) in the first three seconds after photoionization yields a photon scatter rate, which starts immediately at its saturation value and hence indicates a cold ion (black curve).
\vspace{0.3cm}

\section{Conclusion}
In conclusion, we have demonstrated laser spectroscopy, photoionization and laser cooling of a single Yb$^+$ ion on the 4f$^{14}$6s $^2$S$_{1/2}$ - 4f$^{13}$5d6s $^3$D$[3/2]_{1/2}$ transition. We measure the isotope shifts and determine the specific mass shift as well as the field shifts. The results on loading-and-cooling show that this transition complements the existing manipulation capabilities  of Yb$^+$ ions and potentially could replace the $^2$S$_{1/2}$-$^2$P$_{1/2}$ transition. In the future, we plan to make use of the relatively narrow line width of this transition to facilitate Doppler cooling to very low temperatures, potentially directly into the vibrational ground state. Moreover, the new availability of a $\Lambda$-system in Yb$^+$ will allow for new ways of quantum state manipulation and, for example, EIT cooling \cite{Morigi2000,Roos2000}.

\section{Acknowledgments}
We would like to thank Sebastien Garcia, Neil \mbox{McDonald}, Ben Metcalf, Christopher Overstreet and the workshop of the
Cavendish Laboratory. This work has been supported by EPSRC (EP/H005679/1) and ERC (Grant No. 240335).

\end{document}